\documentclass[prl,longbibliography,twocolumn,superscriptaddress,amsmath,amssymb,verbatim]{revtex4-1}

\usepackage{graphicx}
\usepackage{lmodern}
\usepackage{epstopdf}
\usepackage[colorlinks=true,citecolor=blue,linkcolor=blue,urlcolor=blue,bookmarks=true]{hyperref}
\usepackage{xcolor}
\definecolor{green}{rgb}{0.1, 0.6, 0.3} 
\newcommand{\red}{\color{black}}

\newcommand{\YCu}{YCu$_3$(OH)$_6$Cl$_3$}

\begin{document}

\preprint{APS/123-QED}

\title{Negative-vector-chirality 120$^\circ$ spin structure in the defect- and distortion-free quantum kagome antiferromagnet YCu$_3$(OH)$_6$Cl$_3$}


\author{A. Zorko}
\email{andrej.zorko@ijs.si}
\affiliation{Jo\v{z}ef Stefan Institute, Jamova c.~39, SI-1000 Ljubljana, Slovenia}
\author{M. Pregelj}
\affiliation{Jo\v{z}ef Stefan Institute, Jamova c.~39, SI-1000 Ljubljana, Slovenia}
\author{M.\,Gomil\v{s}ek}
\affiliation{Jo\v{z}ef Stefan Institute, Jamova c.~39, SI-1000 Ljubljana, Slovenia}
\affiliation{Centre for Materials Physics, Durham University, South Road, Durham, DH1 3LE, UK}
\author{M. Klanj\v{s}ek}
\affiliation{Jo\v{z}ef Stefan Institute, Jamova c.~39, SI-1000 Ljubljana, Slovenia}
\author{O. Zaharko}
\affiliation{Laboratory for Neutron Scattering, Paul Scherrer Institute, CH-5232 Villigen PSI, Switzerland}
\author{W. Sun}
\affiliation{Fujian Provincial Key Laboratory of Advanced Materials, Department of Materials Science and Engineering, College of Materials, Xiamen University, Xiamen 361005, Fujian Province, People's Republic of China}
\author{J.-X. Mi}
\affiliation{Fujian Provincial Key Laboratory of Advanced Materials, Department of Materials Science and Engineering, College of Materials, Xiamen University, Xiamen 361005, Fujian Province, People's Republic of China}

\date{\today}

\begin{abstract}
The magnetic ground state of the ideal quantum kagome antiferromagnet (QKA) has been a long-standing puzzle, mainly because perturbations to the nearest-neighbor isotropic  Heisenberg Hamiltonian can lead to various fundamentally different ground states. 
Here we investigate a recently synthesized QKA representative {\YCu}, where perturbations commonly present in real materials, like lattice distortion and intersite ion mixing, are absent.
Nevertheless, this compound enters a long-range magnetically ordered state below $T_N=15$\,K.
Our powder neutron diffraction experiment reveals that its magnetic structure corresponds to a coplanar $120^\circ$ state with negative vector spin chirality.
The ordered magnetic moments are suppressed to $0.42(2)\mu_B$, which is consistent with the previously detected spin dynamics persisting to the lowest experimentally accessible temperatures. 
This indicates either a coexistence of magnetic order and disorder or the presence of strong quantum fluctuations in the ground state of {\YCu}.
{\red The origin of the magnetic order is sought in terms of Dzyaloshinskii-Moriya magnetic anisotropy and further-neighbor isotropic exchange interactions.}
\end{abstract}

\maketitle

\section{Introduction}
 
\begin{figure}[hb]
\includegraphics[trim = 2mm 0mm 2mm 0mm, clip, width=0.9\linewidth]{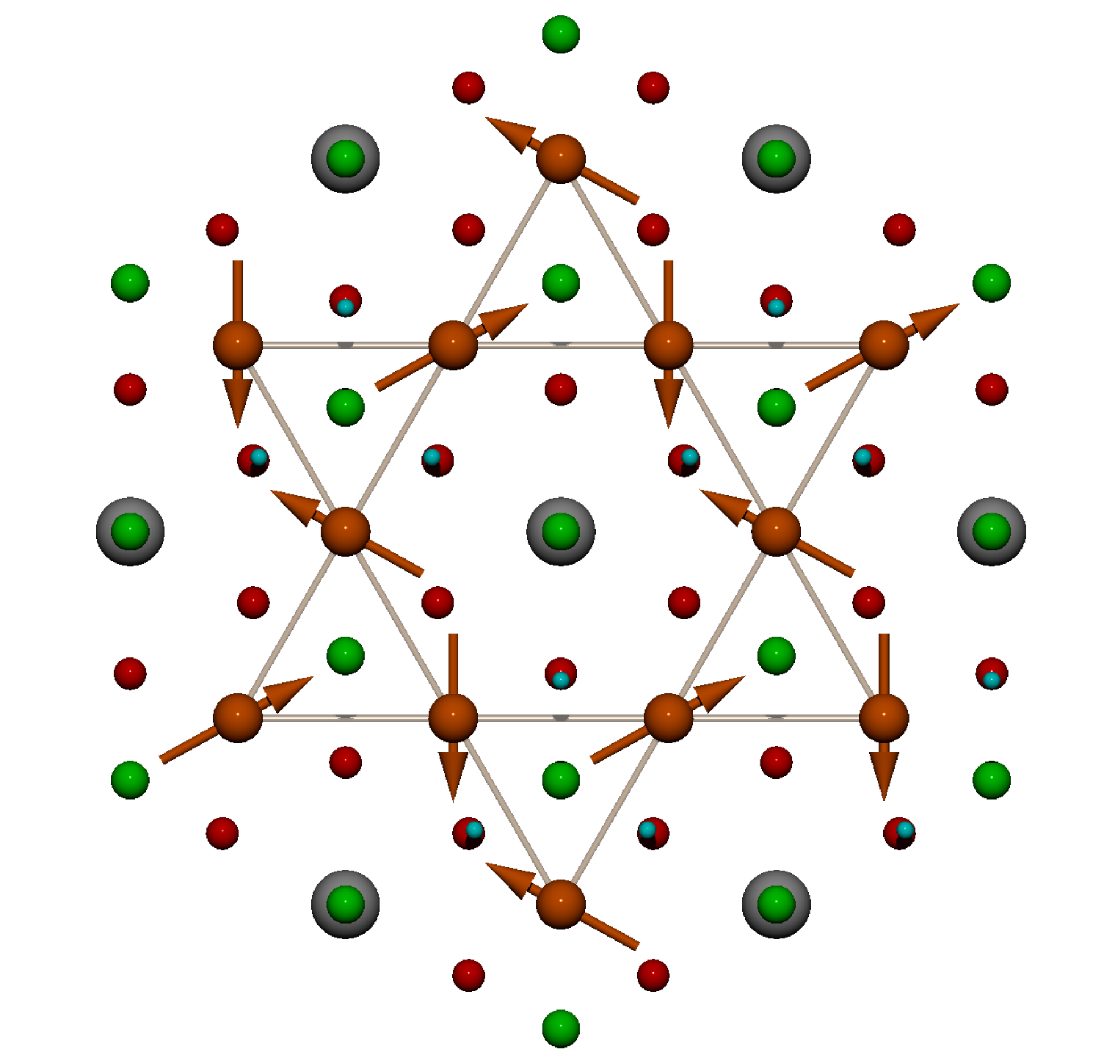}
\caption{A perfect kagome lattice (lines) of Cu$^{2+}$ spin-1/2 ions (orange) is established in the $ab$ plane of the \YCu~compound. 
The Y$^{3+}$, O$^{2-}$, H$^+$, and Cl$^{-}$ ions are shown in gray, red, turquoise, and green, respectively.
A coplanar magnetic structure (arrows) with negative vector spin chirality corresponding to the magnetic propagation vector ${\bf k}=(0,0,0.5)$ and an average magnetic moment of $\mu=0.42(2)\,\mu_B$ per Cu$^{2+}$ site is established below $T_N=15$~K.}
\label{fig1}
\end{figure}

The two-dimensional quantum kagome antiferromagnet (QKA) is a paradigm of geometrical frustration.
It has been intensively studied in the last two decades, mainly because the properties of its predicted spin-liquid ground state turned out to be extremely puzzling \cite{balents2010spin, savary2017quantum, zhou2017quantum}.
Due to a lack of proper approximations it remains unclear, even in theory, whether this state should be gapped or gapless \cite{yan2011spin, depenbrock2012nature,iqbal2013gapless,he2017signatures,liao2017gapless,mei2017gapped}. 
Moreover, various perturbations to the nearest-neighbor isotropic Heisenberg Hamiltonian, like structural distortion and disorder-induced randomness \cite{kawamura2014quantum,shimokawa2015static,rousochatzakis2009dzyaloshinskii}, inter-layer \cite{gotze2016route} and further-neighbor \cite{messio2011lattice,messio2012kagome,suttner2014renormalization,
iqbal2015paramagnetism,gong2015global,buessen2016competing,he2015distinct,
hering2017functional} exchange interaction, as well as magnetic anisotropy \cite{rousochatzakis2009dzyaloshinskii,elhajal2002symmetry,cepas2008quantum,
chernyshev2014quantum,he2015distinct,essafi2016kagome,essafi2017generic,hering2017functional, lee2018gapless,changlani2018macroscopically,zhu2019identifying} may all play an important role in stabilizing very different ground states, from spin liquids and valence-bond solids to magnetically ordered states.
Experimentally, these perturbations usually cannot be controlled in a given material and are rather poorly determined \cite{norman2016herbertsmithite}.
Therefore, experimental realizations of the QKA model, where such perturbations are minimized and limited in number are of paramount importance. 

\begin{table*}[tb]
 \centering
 \caption{Fractional atomic coordinates and lattice parameters of YCu$_3$(OH)$_6$Cl$_3$ measured at 20 K within the space group $P\bar{3}m1$ (No.\,164). The refinement was performed using the FullProf Suite \cite{rodriguez1993recent}. Lattice constants are $a=6.7474(1)$\,\AA, $c = 5.5905(1)$\,\AA, $\alpha$ = $\beta$ = 90$^\circ$, $\gamma$ = 120$^\circ$.}
  \begin{tabular*}{\linewidth}{@{\extracolsep{\fill}} c c c c c c c }
   \hline \hline
Atom & Wyckoff & Site &    $x$     &     $y$     &     $z$     &     Occupancy    \\
\hline
Y11 & $1b$ &  $\bar{3}m.$   &  0.0000   &  0.0000    &  0.5000   & 0.95(5) \\
Y12 & $2c$ &  $3m.$           &  0.0000   &  0.0000    &  0.378 (fixed)   & 0.05(5) \\
Cu   & $3f$ &  $.2/m.$         &  0.5000   &  0.5000    &  0.5000   & 1        \\
Cl1  & $2d$ &  $3m.$           &  0.3333   &  0.6667    &  0.8668(5) & 1        \\
Cl2  & $1a$ &  $\bar{3}m.$   &  0.0000   &  0.0000    &  0.0000   & 1        \\
O    & $6i$  &  $.m.$            &  0.1899(9) &  0.3899(18) &  0.3624(3) & 1       \\
H    & $6i$  &  $.m.$            &  0.2058(8) &  0.4118(16) &  0.1920(6) & 1       \\
   \hline \hline
 \end{tabular*}
 \label{tab:crystal}
\end{table*}

The recently synthesized {\YCu} compound \cite{sun2016perfect} features a perfect-symmetry kagome lattice of Cu$^{2+}$ ($S=1/2$) ions (Fig.\,\ref{fig1}).
This avoids the issue of reduced symmetry present in some QKA representatives, including the paradigmatic herbertsmithite \cite{norman2016herbertsmithite}, where a subtle lattice distortion was found at low temperatures \cite{zorko2017symmetry}. 
Most QKA representatives, including herbertsmithite, also suffer from strong intersite ion disorder leading to the famous QKA impurity problem \cite{norman2016herbertsmithite,gomilsek2016muSR}.
Due to the very large size of the Y$^{3+}$ ions there is no detectable intersite ion disorder present in {\YCu} \cite{sun2016perfect}. 
Hence, because of its much lower level of perturbations to the ideal isotropic Heisenberg Hamiltonian, {\YCu} was initially expected to provide novel insight into the pressing issue of the spin-liquid ground state of the QKA model.
Indeed, after the first bulk magnetic characterizations this system was suggested as a new promising spin-liquid candidate, since no strong anomalies were detected in either its magnetization \cite{sun2016perfect}, or in its heat capacity \cite{puphal2017strong} down to the lowest accessible temperatures despite its sizable Weiss temperature of $-99$\,K \cite{sun2016perfect}.

However, more sensitive local-probe muon spin relaxation ($\mu$SR) measurements have recently disclosed static internal magnetic fields that develop in {\YCu} below $T_N=15$\,K \cite{zorko2019YCu3muon, berthelemy2019local}.
The magnetic ordering at $T_N$ is, in fact also witnessed in an increase of the bulk magnetization below $T_N$ and in heat capacity as a broad maximum at $T_N$ \cite{zorko2019YCu3muon}.
However, the order appears to be rather unconventional, as it progressively sets in and is fully established over the whole sample only for $T\lesssim T_N /3$ \cite{zorko2019YCu3muon}. 
Moreover, persistent spin dynamics was detected by $\mu$SR at temperatures as low as $T/T_N=1/300$ \cite{zorko2019YCu3muon}.
This might have various origins, including emergent spin excitations of correlated spin-loop structures \cite{yaouanc2015evidence}, or fragmentation of magnetic moments into an ordered and a fluctuating part, e.g, as proposed for some partially ordered magnetic states on the kagome lattice \cite{essafi2017generic} and for incommensurate ordered states \cite{pregelj2012persistent}.

The fundamental question of why {\YCu} has a magnetically ordered ground state instead of a spin liquid needs to be addressed, even more so because the number of relevant perturbations to the isotropic nearest-neighbor Heisenberg Hamiltonian is substantially reduced in this compound, making it closer to the ideal QKA than most of the known spin-liquid QKA representatives.
In order to be able to address this important question, a proper characterization of the magnetic order is required, which calls for complementary experiments, like neutron diffraction.
We note that an earlier neutron diffraction experiment failed to detect any magnetic Bragg peaks \cite{berthelemy2019local}.
It needs to be stressed though, that the experiment was focused on structural refinement and was, therefore, quite coarse in resolution.
Here we report the results of a neutron diffraction study with much better resolution and statistics, which provides clear evidence for the appearance of magnetic Bragg peaks below $T_N$. 
These are assigned to a uniform 120$^\circ$ magnetic order with negative vector spin chirality (Fig.\,\ref{fig1}), the magnetic propagation vector ${\bf k}=(0,0,0.5)$ and an average ordered magnetic moment of $\mu=0.42(2)\mu_B$ at 1.5\,K.
The origin of this magnetic order is discussed in terms of magnetic anisotropy and exchange interactions beyond the nearest-neighbor Heisenberg exchange.

\section{Experimental Details}
Powder neutron diffraction was performed on the DMC powder diffractometer at the Paul Scherrer Institute, Villigen, Switzerland.
2.3\,g of sample was put in an Al sample can measuring 6\,mm in diameter.
The measurements were performed at 1.5\,K and 20\,K at a fixed incoming-neutron energy corresponding to a wave length of $\lambda= 4.506$\,\AA.
The high background in the neutron diffraction patterns is due to strong incoherent scattering from the hydrogen nuclei present in the sample.
In order to detect weak magnetic Bragg reflections, high-statistics runs were collected, with the measurement time of 60\,h at each temperature.
All measurements were performed on a sample from the same batch as the one used in our previous $\mu$SR investigation \cite{zorko2019YCu3muon}.
\begin{figure}[t]
\includegraphics[trim = 0mm 2mm 0mm 2mm, clip, width=1\linewidth]{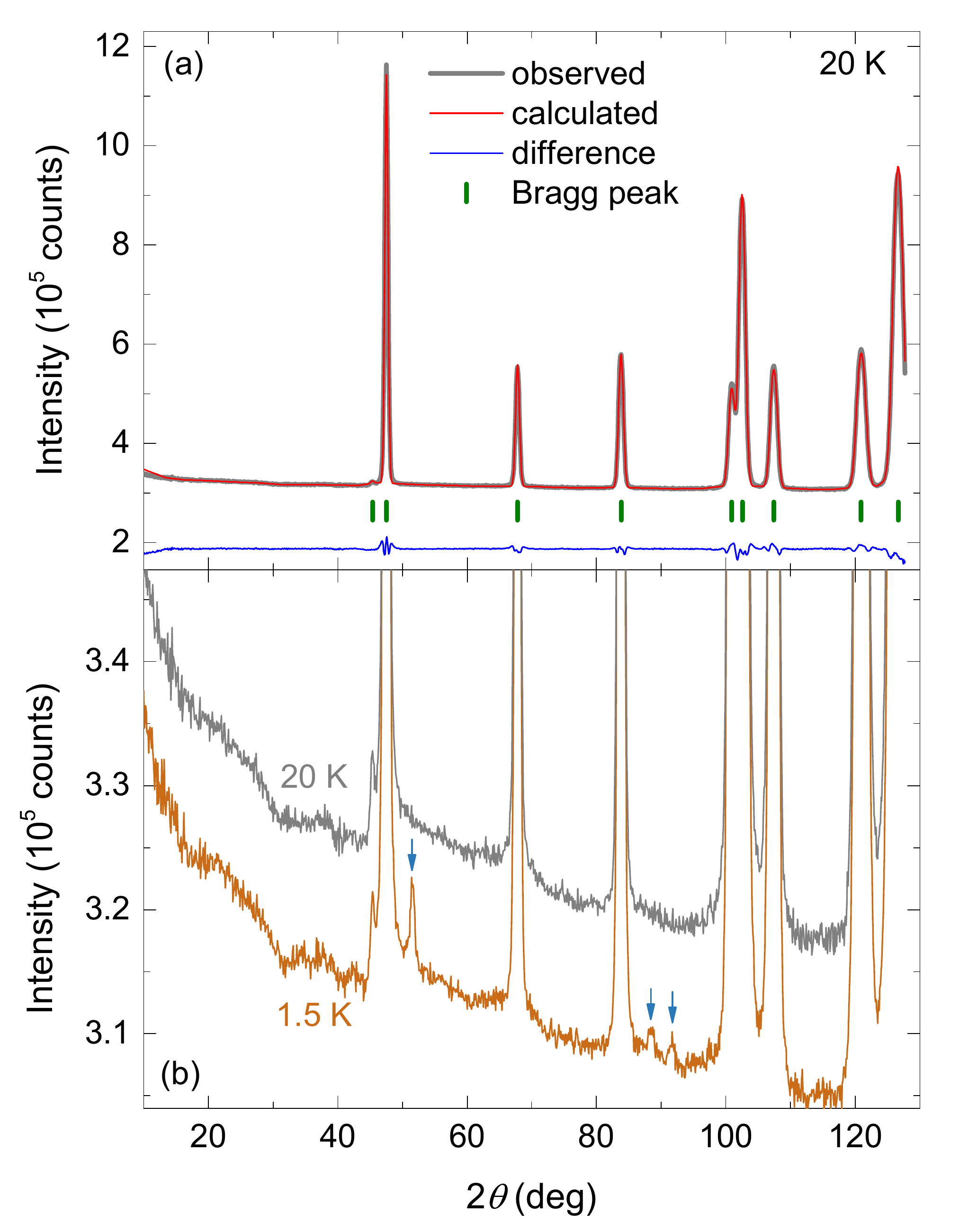}
\caption{(a) The neutron-diffraction pattern of \YCu taken at 20\,K. 
The calculation is based on the $P\bar{3}m1$ space group with trigonal symmetry that accounts well for the crystal structure.
The difference between the observed and the calculated diffraction patterns is displaced vertically for clarity.
(b) The additional (magnetic) Bragg peaks that appear below $T_N=15$\,K are marked by arrows in the dataset taken at 1.5\,K.
Here, the dataset taken at 20\,K is translated vertically for clarity.
}
\label{fig2}
\end{figure}

\section{Results}

The neutron diffraction pattern above $T_N$ (at 20\,K) can be perfectly reproduced using the trigonal crystal-structure model with the $P\bar{3}m1$ symmetry (space group No.\,164) initially found by Sun {\it et al.}~using X-ray diffraction (XRD) \cite{sun2016perfect}.
Due to high sensitivity of neutrons to light nuclei, the position of the hydrogen atoms is better determined here than in the previous XRD study.
The good agreement between the model and our measurements shown in Fig.\,\ref{fig2}(a) confirms the high quality of our sample.
The refined parameters of the crystal structure are summarized in Table\,\ref{tab:crystal}.
They are consistent with the previously published XRD results at higher temperature \cite{sun2016perfect} and the recently published neutron-diffraction results obtained at 1.6\,K \cite{berthelemy2019local}.
The precision of some of the derived crystal-structure parameters is not very high, as the focus of our study was on the magnetic order, which is why a large wavelength of the incoming neutrons was chosen. 
This allowed for better sensitivity at low $q$'s where weak magnetic reflections were expected to appear.  
As a result, the occupancy of the sites Y11 and Y12 have significant uncertainties.
{\red In fact, refinement with a single Y site at the Wyckoff position $1b$ yields only a slightly worse $R_F$ factor of 0.79 compared to the value of 0.76 obtained for two sites.  
This is consistent with recent neutron diffraction experiment that suggested the presence of a single Y site \cite{berthelemy2019local}.}

\begin{table*}[t]
 \centering
 \caption{The basis vectors $\psi_i^j$ of the irreducible representations (irreps) $\Gamma_2$, $\Gamma_4$ and $\Gamma_5$ of the space group $P\bar{3}m1$ for the magnetic propagation vector $\mathbf{k}=(0,0,0.5)$ occurring in the magnetic representation at the magnetic site $(1/2, 1/2, 1/2)$ of atom 1 and its two crystallographically equivalent sites. The representation analysis was performed using the BasIreps program incorporated in the FullProf Suite \cite{rodriguez1993recent}. The listed basis vectors are scaled to yield normalized magnetic moments in the complex plane.}
  \begin{tabular*}{\linewidth}{@{\extracolsep{\fill}} c  c  c c c  c  c c c  c  c c c }
   \hline \hline
 irrep & Basis &  \multicolumn{3}{c}{Atom 1} & & \multicolumn{3}{c}{Atom 2} & & \multicolumn{3}{c}{Atom 3} \\
& vector & $m_a$ & $m_b$ & $m_c$ & & $m_a$ & $m_b$ & $m_c$ & & $m_a$ & $m_b$ & $m_c$\\ 
\hline
$\Gamma_2$ & $\psi_2^1$ & $\frac{1}{\sqrt{3}}$ & $-\frac{1}{\sqrt{3}}$ & 0 & & $\frac{1}{\sqrt{3}}$ &  $\frac{2}{\sqrt{3}}$ & 0 & & $-\frac{2}{\sqrt{3}}$ &  $-\frac{1}{\sqrt{3}}$ & 0 \\
                   & $\psi_2^2$ & 0 &  0 & 1 & & 0 & 0 & 1 & & 0 & 0 & 1 \\
\hline
$\Gamma_4$ & $\psi_4^1$ & 1 & 1 & 0 & & $-1$ & 0 & 0 & & 0 & $-1$ & 0 \\
\hline
$\Gamma_5$ & $\psi_5^1$ & 1 & 0 & 0 & & 0 & $-\frac{1}{2} - i \frac{\sqrt{3}}{2}$ & 0 & & $\frac{1}{2} - i \frac{\sqrt{3}}{2}$ & $\frac{1}{2} - i \frac{\sqrt{3}}{2}$ & 0 \\
                   & $\psi_5^2$ & 0 & 1 & 0 & & $\frac{1}{2} + i \frac{\sqrt{3}}{2}$ & $\frac{1}{2} + i \frac{\sqrt{3}}{2}$ & 0 & & $-\frac{1}{2} + i \frac{\sqrt{3}}{2}$ & 0 & 0 \\
                   & $\psi_5^3$ & 0 & 0 & 1 & & 0 & 0 & $-\frac{1}{2} - i \frac{\sqrt{3}}{2}$ & & 0 & 0 & $-\frac{1}{2} + i \frac{\sqrt{3}}{2}$ \\
                   & $\psi_5^4$ & 0 & $-1$ & 0 & & $-\frac{1}{2} + i \frac{\sqrt{3}}{2}$ & $-\frac{1}{2} + i \frac{\sqrt{3}}{2}$ & 0 & & $\frac{1}{2} + i \frac{\sqrt{3}}{2}$ & 0 & 0 \\
                   & $\psi_5^5$ & $-1$ & 0 & 0 & & 0 & $\frac{1}{2} - i \frac{\sqrt{3}}{2}$ & 0 & & $-\frac{1}{2} - i \frac{\sqrt{3}}{2}$ & $-\frac{1}{2} - i \frac{\sqrt{3}}{2}$ & 0 \\
                   & $\psi_5^6$ & 0 & 0 & 1 & & 0 & 0 & $-\frac{1}{2} + i \frac{\sqrt{3}}{2}$ & & 0 & 0 & $-\frac{1}{2} - i \frac{\sqrt{3}}{2}$ \\
   \hline \hline
 \end{tabular*}
 \label{tab:irreps}
\end{table*}
\begin{figure*}[t]
\includegraphics[trim = 0mm 0mm 0mm 0mm, clip, width=1\linewidth]{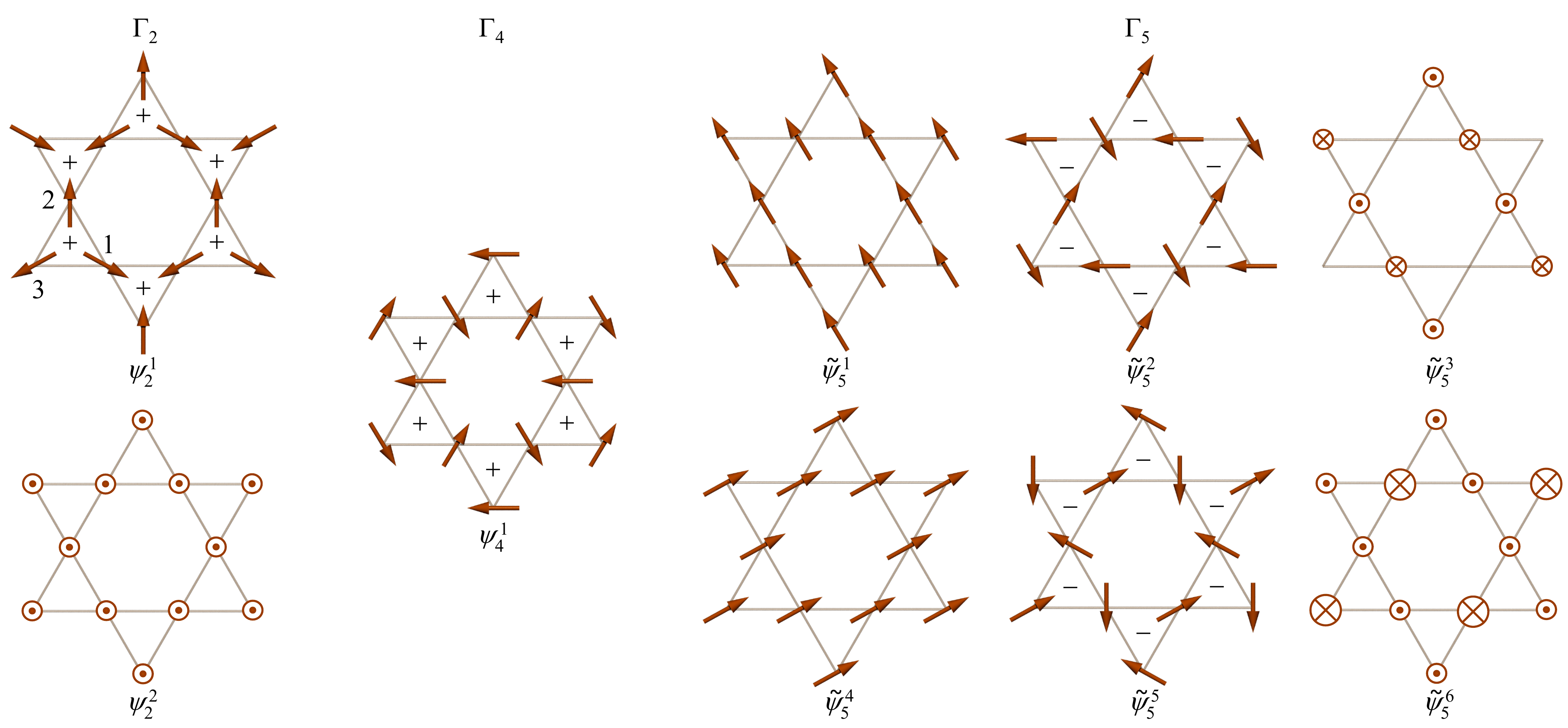}
\caption{The spin structures corresponding to the basis vectors of the irreducible representations $\Gamma_i$ for the magnetic propagation vector $\mathbf{k}=(0,0,0.5)$ occurring at the magnetic Cu$^{2+}$ site in {\YCu}. 
For $\Gamma_2$ and $\Gamma_4$ the basis vectors are given in Table\,\ref{tab:irreps}, while the six presented spin structures of $\Gamma_5$ correspond to the linear combinations
$\tilde{\psi_5^1}$ = $iA \psi_5^1$ + $B_{++} \psi_5^2$ + $B_{-+} \psi_5^4$ + $iA \psi_5^5$,
$\tilde{\psi_5^2}$ = $B_{--} \psi_5^1$ $-$ $iA \psi_5^2$ $-$ $iA \psi_5^4$ + $B_{+-} \psi_5^5$,
$\tilde{\psi_5^3}$ = $C_{+} \psi_5^3$ + $C_{-} \psi_5^6$,
$\tilde{\psi_5^4}$ = $A \psi_5^1$ + $iB_{--} \psi_5^2$ + $iB_{-+} \psi_5^4$ $-$ $A \psi_5^5$,
$\tilde{\psi_5^5}$ = $iB_{++} \psi_5^1$ $-$ $A \psi_5^2$ + $A \psi_5^4$ + $iB_{+-} \psi_5^5$, and
$\tilde{\psi_5^6}$ = $D_{+} \psi_5^3$ + $D_{-} \psi_5^6$, where
$A$\,=$1/\sqrt{3}$, $B_{\pm\pm}$\,=\,$(\pm1\pm i/\sqrt{3})/2$,
$C_{\pm}$\,=\,$(-\sqrt{3}\pm i)/4$, and $D_{\pm}$\,=\,$(1\pm i/\sqrt{3})/\sqrt{8}$,
which represent a real basis of $\Gamma_5$ \cite{essafi2017generic}.
The chosen assignment of the three magnetically nonequivalent sites in Eq.\,(\ref{chirality}) is shown for $\psi_2^1$. 
The spin structures $\psi_2^1$ and $\psi_4^1$ have positive vector chirality +1 [see Eq.\,(\ref{chirality})] for each triangle when projected onto the $c$ axis, while this projection is -1 for the $\tilde\psi_5^2$ and $\tilde\psi_5^5$ states.
These chirality vectors are perpendicular to the kagome planes and are denoted by $\pm$ signs inside the triangles.
The vector chirality of all other structures is zero as they are collinear. 
}
\label{fig3}
\end{figure*}
The diffraction pattern below $T_N$ (at 1.5\,K) reveals the emergence of additional weak Bragg reflections, highlighted by arrows in Fig.\,\ref{fig2}(b), while the dominant reflections already present above $T_N$ are unchanged.
As there is no structural transition in {\YCu} at least down to 1.6\,K \cite{berthelemy2019local}, these new reflections are a fingerprint of long-range magnetic order appearing below $T_N$.
The positions of the magnetic reflections correspond to the magnetic propagation vector $\mathbf{k}=(0,0,0.5)$, signifying a $q=0$ type magnetic structure \cite{messio2011lattice} within the $ab$ kagome planes and the antiferromagnetic nature of the magnetic order between neighboring kagome planes.
The comparable width of the magnetic and the structural Bragg reflections reveals that the correlation length of the magnetic order is very large.

We tackle the problem of determining the magnetic structure in {\YCu} within the framework of representation analysis \cite{wills2001long}, using the FullProf Suite \cite{rodriguez1993recent}. 
This analysis reveals that the little group for the propagation vector $\mathbf{k}=(0,0,0.5)$ has six irreducible representations (irreps) $\Gamma_i$.
From these only three occur at the Cu$^{2+}$ crystallographic site (Wyckoff position $3f$), where the magnetic representation $\Gamma$ is decomposed as
\begin{equation}
\label{irrdecomp}
\Gamma=0\Gamma_1+2\Gamma_2+0\Gamma_3+1\Gamma_4+3\Gamma_5+0\Gamma_6.
\end{equation}
According to Landau's theory of second-order phase transition, the magnetic structure can correspond to only one of the three non-zero irreps  \cite{wills2001long}: either the one-dimensional $\Gamma_2$ or $\Gamma_4$, or the two-dimensional $\Gamma_5$. 
The normalized basis vectors $\psi_i^j$ for these representations are given in Table\,\ref{tab:irreps}.
Here the index $i$ corresponds to a specific irrep while the index $j$ counts its $d_i\cdot n_i$ basis vectors, where $d_i$ is the dimensionality of $\Gamma_i$ and $n_i$ its multiplicity in the decomposition given by Eq.~(\ref{irrdecomp}).
The corresponding spin structures are shown in Fig.\,\ref{fig3}.
For $\Gamma_5$ where the basis vectors are complex and three pairs related by complex conjugation can be formed, we rather show their six linear combinations $\tilde{\psi}_5^j=\sum_ka_k\psi_5^k$ that constitute an equivalent real basis of this irrep \cite{essafi2017generic}.
The spins on each site are normalized for all basis vectors, except for the antiferromagnetic vectors $\tilde{\psi_5^3}$ and $\tilde{\psi_5^6}$, because it is impossible to form a net zero-magnetization state for three normalized  collinear spins.
The basis vectors $\psi_2^1$, $\psi_4^1$ and $\tilde \psi_5^2$, $\tilde \psi_5^5$ correspond to non-collinear structures within the kagome plane, the so-called 120$^\circ$ states, which satisfy the condition $\sum_i {\bf S}_i=0$ on each spin triangle. 
The other basis vectors describe either ferro-, ferri- or antiferromagnetic collinear structures.
The coplanar 120$^\circ$ structures can be further distinguished based on their vector chirality 
\begin{equation}
\kappa = \frac{2}{3\sqrt{3}S^2}({\bf S}_1\times {\bf S}_2 + {\bf S}_2\times {\bf S}_3 + {\bf S}_3\times {\bf S}_1).
\label{chirality}
\end{equation}
The projection of the vector chirality on the $c$ axis is $+1$ for $\psi_2^1$ and $\psi_4^1$, while it is $-1$  for $\tilde \psi_5^2$ and $\tilde \psi_5^5$.
The sign of these projections is determined by the anticlockwise site assignment shown in  Fig.\,\ref{fig3}.
Although the two positive-chirality structures are related by a global 90$^\circ$ rotation of spins, the representations $\Gamma_2$ and $\Gamma_4$ differ in that an out-of plane ferromagnetic component can be added within the former but not within the  latter representation. This gives rise to the so-called umbrella structure for $\Gamma_2$ \cite{wills2001long}. 

\begin{table}[t]
 \centering
 \caption{The in-plane ($\mu_{ab}$) and the out-of-plane ($\mu_c$) components of the ordered magnetic moment $\mu$ for the different spin structures corresponding to the allowed irreducible representations (irreps). The quality of the refinement for each structure is given by the reduced $\chi^2$ value.}
  \begin{tabular*}{\linewidth}{@{\extracolsep{\fill}} c  c c c c c  c}
   \hline \hline
irrep & Vector & Atom & $\mu_{ab}/\mu_B$ & $\mu_c/\mu_B$ & $\mu/\mu_B$ & $\chi^2$\\ 
\hline
$\Gamma_2$ & $\psi_2^1$ & 1--3 & 0.48(2) & 0 & 0.48(2) & 2.57 \\
\hline
$\Gamma_4$ & $\psi_4^1$ & 1--3 & 0.33(2) & 0 & 0.33(2) & 1.61\\
\hline
$\Gamma_5$ & $\tilde{\psi}_5^5$ & 1--3 & 0.42(2) & 0 & 0.42(2) & 1.23\\
 & $\tilde{\psi}_5^6$ & 1,\,2 & 0 & 0.26(1) & 0.26(1) & 1.40\\
 & & 3 & 0 & -0.52(2) & 0.52(2) & \\
 & $0.68\tilde{\psi}_5^5+0.32\tilde{\psi}_5^6$ & 1,\,2 & 0.36(4) & 0.14(4) & 0.39(5) & 1.22\\
 & & 3 & 0.36(4) & -0.27(7) & 0.46(6) & \\
   \hline \hline
 \end{tabular*}
 \label{tab:results}
\end{table}

\begin{figure}[t]
\includegraphics[trim = 0mm 0mm 0mm 0mm, clip, width=1\linewidth]{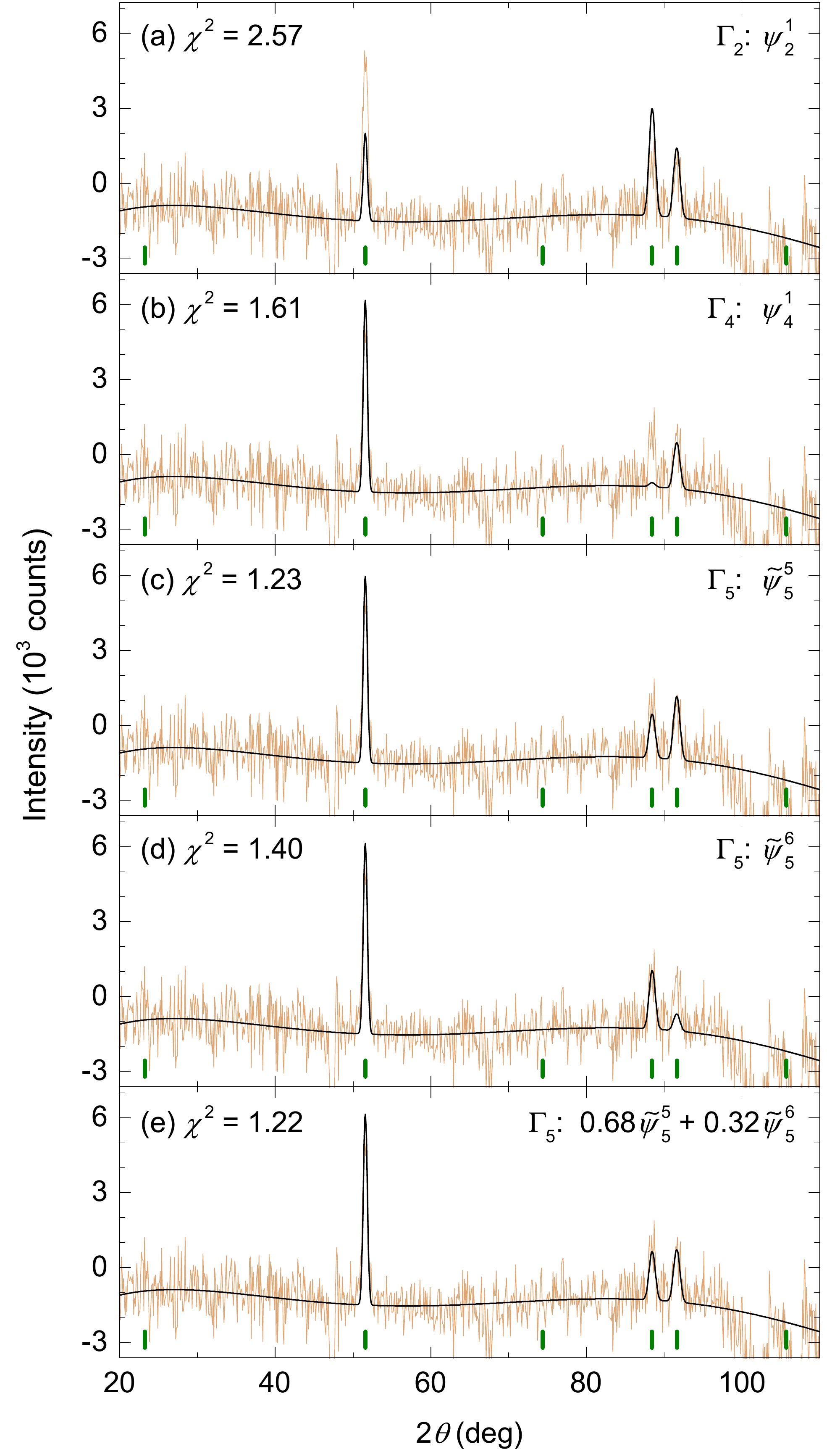}
\caption{Simulations of the experimental magnetic diffraction pattern obtained by subtracting the dataset taken at 20\,K from the dataset taken at 1.5\,K, which conform to models based on different irreducible representations $\Gamma_i$ of the little group corresponding to the magnetic propagation vector ${\bf k}=(0,0,0.5)$.
The positions of all the predicted magnetic reflections for this ${\bf k}$ are marked by green ticks.
{\red The intensities of the reflections at $23^\circ$, $74^\circ$, and $105^\circ$ are non-zero only for ferromagnetic spin structures.}
The quality of the fit for each model is reflected in the reduced $\chi^2$ values, which are summarized in Table\,\ref{tab:results} together with the corresponding average magnetic moments.
}
\label{fig4}
\end{figure}

The refinement of the magnetic structure model to the magnetic diffraction pattern, i.e., to the difference between the diffraction pattern recorded at 1.5\,K and the one at 20\,K, is shown in Fig.\,\ref{fig4} for all relevant irreps.
The corresponding parameters are summarized in Table\,\ref{tab:results}.
First, we note that our powder neutron-diffraction experiment does not distinguish between the spin structures $\tilde{\psi}_5^1$ and $\tilde{\psi}_5^4$, $\tilde{\psi}_5^2$ and $\tilde{\psi}_5^5$, and $\tilde{\psi}_5^3$ and $\tilde{\psi}_5^6$.
Therefore, we can focus on one of them for each pair when considering the irrep $\Gamma_5$.
For the irrep $\Gamma_2$ the agreement is worse than for the other two allowed irreps, giving by far the largest reduced $\chi^2$ value after refinement [Fig.\,\ref{fig4}(a)]. 
In this solution only the in-plane component $\psi_2^1$ is included, because the ferromagnetic out-of-plane component $\psi_2^2$ gives a strong reflection at $105^\circ$, which is not experimentally observed.
For the other one-dimensional irrep $\Gamma_4$, also yielding a coplanar spin structure with positive vector chirality, the agreement with experiment is somewhat better, although it severely underestimates the intensity of the Bragg peak at $88^\circ$ [Fig.\,\ref{fig4}(b)].
A much better agreement is obtained for the coplanar spin structure with negative vector chirality $\tilde \psi_5^5$ from the two-dimensional $\Gamma_5$ [Fig.\,\ref{fig4}(c)].
On the other hand, for this irrep, the in-plane ferromagnetic spin structure $\tilde \psi_5^4$ does not fit the experiment well, as it yields a strong reflection at 23$^\circ$, which is absent in experiment.
However, the collinear out-of-plane spin structure $\tilde \psi_5^6$ is consistent with the observed magnetic Bragg peaks [Fig.\,\ref{fig4}(d)], even though it gives a somewhat worse agreement with experiment than the $\tilde \psi_5^5$ spin structure.
We note than a marginally better agreement than for the pure coplanar spin structure $\tilde \psi_5^5$ is obtained for a linear combination of the basis vectors $0.68\tilde \psi_5^5+0.32\tilde \psi_5^6$. 
Within this best-fit model, the relative uncertainty of the two mixing coefficients is, however, quite big (15\%).
Consequently, the uncertainty of the ordered moments within this model is also increased compared to the pure $\tilde \psi_5^5$ model (see Table\,\ref{tab:results}). 
Independent of the model, we find that the magnitude of the ordered moments is significantly reduced from the full value of $1-1.2\mu_B$ expected for Cu$^{2+}$.

\section{Discussion}

Our refinements yield two candidate magnetic structures that are almost indistinguishable regarding the quality of the fit (Fig.\,\ref{fig4}): the in-plane coplanar 120$^\circ$ spin structure with negative vector chirality given by the basis vector $\tilde{\psi}_5^5$ and the spin structure represented by a linear combination of two basis vectors from $\Gamma_5$, $0.68\tilde \psi_5^5+0.32\tilde \psi_5^6$, where the collinear out-of-plane component $\tilde{\psi}_5^6$ is added to the same in-plane coplanar component.
These two spin structures can, however, be distinguished based on their energy.
The isotropic Heisenberg exchange model with dominant antiferromagnetic interactions prefers either coplanar or more complex spin structures with larger magnetic unit cells in the kagome plane, the latter if interactions beyond the nearest neighbors are considered \cite{messio2011lattice,suttner2014renormalization,iqbal2015paramagnetism,
gong2015global}.
In {\YCu} the in-plane component of the magnetic propagation vector is zero, therefore, the basic triangular unit is representative for energy considerations.  

The distinction between the coplanar structures with positive and negative vector chirality is made by magnetic anisotropy. 
The antisymmetric Dzyaloshinshii-Moriya (DM) anisotropy ${\bf D}\cdot{\bf S}_i\times{\bf S}_j$, where ${\bf D}$ is the DM vector, directly couples to the vector spin chirality of Eq.\,(\ref{chirality}). 
The out-of-plane DM component $D_z$ therefore acts as an easy-plane \cite{elhajal2002symmetry} and selects either the state with positive vector chirality for $D_z<0$ or the state with negative vector chirality for $D_z>0$, the latter apparently being the situation in {\YCu}.
For the former state, the in-plane DM component $D_{p}$ causes out-of-plane canting of the spin structure, i.e., an admixture of the $\psi_2^2$ state, leading to the umbrella structure, while the state with negative chirality remains coplanar even for finite $D_{p}$ \cite{elhajal2002symmetry}.
To explain the possible admixture of the antiferromagnetic collinear state within $\Gamma_5$, as suggested by our experiment, symmetric exchange anisotropy $\Delta S_i^z S_j^z$ with $\Delta > 0$ needs to be included \cite{essafi2017generic}, since it acts as an easy-axis anisotropy perpendicular to the kagome plane.
Therefore, we consider the Hamiltonian
\begin{equation}
\mathcal{H}=\sum_\triangle \left(J_1 {\bf S}_i {\bf S}_j + {\bf D}_{ij}\cdot {\bf S}_i \times {\bf S}_j + \Delta S_i^z S_j^z   \right),
\end{equation}
where $J_1$ is the nearest-neighbor isotropic exchange interaction and the sum runs over the three spins in the basic triangular unit.
For an ordered magnetic moment $\mu=g\mu_B\left\langle S \right\rangle$ in the coplanar $\tilde \psi_5^5$ spin structure, where $g$ and $\mu_B$ are the $g$-factor and the Bohr magneton, respectively, while $\left\langle S \right\rangle$ denotes the expectation value of the spin, this Hamiltonian yields the classical energy per spin site of
$E(\tilde \psi_5^5)=-\frac{1}{2}(J_1+\sqrt{3}D_z)(\mu/g\mu_B)^2$.
In the antiferromagnetic spin structure $\tilde \psi_5^6$ the ordered magnetic moment $\mu$ on one of the sites in the triangle is accompanied by two moments $\mu$/2 on the other two sites (see Fig.\,\ref{fig3}) and the classical energy is
$E(\tilde{\psi}_5^6)=-\frac{1}{4}\left(J_1+\Delta\right) (\mu/g\mu_B)^2$. 
Taking into account the magnitude of the in-plane and the out-of plane magnetic moment components for the solutions $\tilde \psi_5^5$ and $0.68\tilde \psi_5^5+0.32\tilde \psi_5^6$ (Table\,\ref{tab:results}), we obtain the corresponding energies
$E(\tilde \psi_5^5)=-0.088\frac{J_1}{g^2}-0.153\frac{D_z}{g^2}$ 
and
$E(0.68\tilde \psi_5^5+0.32\tilde \psi_5^6)=-0.083\frac{J_1}{g^2}-0.112\frac{D_z}{g^2}-0.020\frac{\Delta}{g^2}$. 
Comparing these values, we find that the composite spin structure $0.68\tilde \psi_5^5+0.32\tilde \psi_5^6$ is energetically favorable only if the condition
$\Delta > 0.25 J_1 + 2.0 D_z$ is satisfied.

Both types of exchange anisotropies are spin-orbit-coupling induced relativistic corrections to the isotropic Heisenberg Hamiltonian.
Their expected values are of the order $D/J_1 \approx \Delta g/g$ and $\Delta/J_1 \approx (\Delta g/g)^2$ \cite{moriya1960anisotropic}.
For the Cu$^{2+}$ ions the $g$-factor anisotropy is of the order $\Delta g/g \approx 0.15$ \cite{abragam1970electron}.
The DM anisotropy is therefore usually dominant for Cu-based kagome lattices \cite{zorko2008dzyaloshinsky,zorko2013dzyaloshinsky}.
We thus conclude that the required $\Delta/J_1$ ratio for the composite spin structure $0.68\tilde \psi_5^5+0.32\tilde \psi_5^6$ to be stable is at least an order of magnitude larger than can be reasonably expected. 
The pure coplanar spin structure with negative vector chirality $\tilde \psi_5^5$ is, therefore, the most likely ground state of {\YCu}.
We note, that since powder neutron diffraction experiment does not distinguish between this state and the other negative-vector-chirality state $\tilde \psi_5^2$ from the same irrep, any linear combination of the two states is possible.
Only the experiment on a single-crystal sample could exactly determine the spin structure.  

The observed spin structure of {\YCu} corresponds to the so-called $q=0$ magnetic order in a given kagome plane, which is one of the regular magnetic orders of this lattice \cite{messio2011lattice}.
In the case of classical spins this state with negative vector spin chirality can be stabilized by an infinitesimal DM interaction $D_z > 0$ \cite{elhajal2002symmetry, hering2017functional}.
However, in the case of quantum spins, the DM interaction needs to surpass a threshold value for the ordered state to replace a spin-liquid ground state if only the nearest-neighbor isotropic exchange interaction is considered.
The majority of theoretical studies suggest that the quantum-critical point between a spin-liquid and the $q=0$ ordered state should occur around $D_z\sim 0.1 J_1$  \cite{cepas2008quantum,rousochatzakis2009dzyaloshinskii,
hering2017functional,zhu2019identifying}.
We note, though, that a much lower critical value of $D_z\sim 0.012(2) J_1$ was recently suggested by a theoretical approach that favors a gapless U(1) spin liquid as the ground state in absence of the DM interaction \cite{lee2018gapless}.

As an alternative to the DM component $D_z$ being responsible for stabilizing the ordered state in {\YCu}, the same state could also be stabilized by exchange interactions beyond nearest neighbors.
A finite next-nearest-neighbor interaction $J_2$, for example, decreases the above-mentioned quantum critical point and stabilizes the $q=0$ ordered state above $J_2/J_1\approx0.15$ \cite{zhu2019identifying}.
Indeed, the $q=0$ ordered state is stable even in the absence of the DM interaction in the parameter range $0.15-0.5 \lesssim J_2/J_1 \lesssim 1.7$ \cite{suttner2014renormalization,gong2015global,hering2017functional}. 
{\red A finite DM component $D_z$ further shifts these boundaries in favor of the ordered state \cite{hering2017functional} and is needed to select the chirality of the spin structure.
The shift of the phase boundaries towards lower $J_2/J_1$ values} is due to the fact that the isotropic Heisenberg antiferromagnetic interaction on the basic triangular unit is more frustrated than the DM interaction -- e.g., in the energy calculation of the coplanar state $\tilde \psi_5^5$ presented above, the DM contribution is effectively multiplied by a factor $\sqrt{3}$ when compared to the $J_1$ contribution.
The DM component $D_z$, therefore, tends to reduce quantum fluctuation in favor of magnetic ordering.
The same is true for the in-plane DM component $D_p$, which also disfavors some spin structures from the ground-state manifold of the classical isotropic Hamiltonian and should be even more efficient than the $D_z$ component in suppressing quantum fluctuations \cite{zorko2013dzyaloshinsky}. 

The Heisenberg interactions with even more distant neighbors further complicate the theoretically predicted phase diagram \cite{messio2011lattice,messio2012kagome,
iqbal2015paramagnetism,gong2015global,buessen2016competing,he2015distinct,
hering2017functional}.
Although these interactions are negligible in the paradigmatic QKA representative herbertsmithite \cite{jeschke2013first}, they turn out to be relevant in some other QKA representatives. 
For instance, in kapellasite where the Zn$^{2+}$ ions occupy the centers of hexagons instead of the inter-layer sites occupied in herbertsmithite, the diagonal exchange interaction across the kagome hexagon in fact dominates over the nearest-neighbor interaction \cite{bernu2013exchange,iqbal2015paramagnetism}.
{\red There, a gapless spin-liquid ground state that hosts unusual non-coplanar dynamical short-range correlations of a so-called cuboc2 regular magnetic order from the classical spin picture was observed \cite{faak2012kapellasite}, which might also be influenced by a severe site-mixing problem on the kagome planes of that material \cite{kermarrec2014spin}.} 
Moreover, for vesignieite it was recently argued that a complex triple-${\bf k}$ octahedral spin structure is stabilized by a dominant antiferromagnetic third-nearest-neighbor exchange \cite{boldrin2018vesignieite}.
On the other hand, the $q=0$ ordered state found in {\YCu} is generally stable in the region of the phase diagram where $J_1$ and $J_2$ interactions dominate \cite{messio2012kagome,gong2015global,buessen2016competing}. 
Therefore, we do not expect exchange interactions beyond the second neighbors to play any significant role in this compound. 
Lastly, we note that magnetic ordering in kagome materials can also be induced by the interlayer exchange coupling \cite{gotze2016route}.
However, the required value of the interlayer exchange is very high, $J_\perp/J_1\gtrsim 0.15$, which is unlikely to be satisfied in {\YCu}, where the neighboring kagome layers are separated by Cl$^-$ ions and the interlayer distance is as large as 5.6\,\AA.    

Finally, our refinement yields an ordered magnetic moment of $\mu=0.42(2)\,\mu_B$ per Cu$^{2+}$ site, which is significantly smaller than the full value of $1-1.2\,\mu_B$ expected for this ion.
The origin of this strong reduction may be two-fold.
Firstly, it can be due to strong quantum renormalization effects, as theoretical predictions suggest that the ordered magnetic moments should be strongly reduced close to the quantum critical point at $D_z/J_1\sim 0.1$ \cite{cepas2008quantum,gong2015global}.
Alternatively, the observed reduction of the average ordered magnetic moment can occur even for classical spins, because some of the order parameters described by the irreps of the kagome lattice correspond to configurations in which spins are not of unit length \cite{essafi2017generic}.
Consequently, there exist extended regions in parameter space where either multiple types of order have to coexist, or partial order that coexists with a magnetically disordered phase is established.
In the latter scenario, magnetic fluctuations arising from an extensive fraction of disordered spin degrees of freedom would persist down to zero temperature. 
This is a possible explanation of the persistent spin dynamics observed in {\YCu} by $\mu$SR experiments \cite{zorko2019YCu3muon}, while the alternative possibility of an incommensurate magnetic order being at the origin of the persistent spin dynamics is ruled out by the magnetic propagation vector ${\bf k}=(0,0,0.5)$ found in our experiment.

{\red In summary, the neutron diffraction results on {\YCu} are consistent with the $\tilde\psi^5_5$  coplanar $q=0$ state with negative vector chirality and strongly reduced magnetic moments.
The other ground-state candidate $0.68\tilde \psi_5^5+0.32\tilde \psi_5^6$ would require unphysically large symmetric exchange anisotropy. 
Based on energy consideration, we conclude that the selection of this particular spin structure is due to the DM magnetic anisotropy.
The origin of the magnetic ordering below $T_N$ is either in the DM anisotropy or in further-neighbor isotropic exchange interactions.} 

\section{Conclusions}
Our powder neutron diffraction experiments on {\YCu} have disclosed clear magnetic Bragg peaks that appear below $T_N=15$\,K.
Their positions and intensities are consistent with a negative-vector-chirality $120^\circ$ coplanar spin structure within the kagome planes with antiferromagnetic order between the neighboring planes, as described by the magnetic propagation vector ${\bf k}=(0,0,0.5)$.
The detected magnetic order is partial in the sense that the ordered magnetic moment amounts to only $\mu=0.42(2)\mu_B$ at 1.5\,K. 
This strong moment reduction may either arise from strong quantum fluctuations due to the vicinity of a quantum critical point \cite{cepas2008quantum}, or from a coexistence of order and disorder, which could be ubiquitous even in classical kagome systems \cite{essafi2017generic}.
This explains the previously observed spin dynamics that persists down to extremely low temperatures \cite{zorko2019YCu3muon}.
Either Dzyaloshinskii-Moriya anisotropy or further-neighbor exchange interactions could be responsible for stabilizing the observed magnetically ordered state instead of a quantum-spin-liquid state.
The selection of a negative-vector-chirality state out of the degenerate manifold of classical states of the isotropic Heisenberg Hamiltonian is attributed to the out-of-plane Dzyaloshinskii-Moriya anisotropy $D_z >0$. 
Additional experiments, e.g., electron spin resonance, which is extremely sensitive to magnetic anisotropy \cite{zorko2008dzyaloshinsky,zorko2013dzyaloshinsky,kermarrec2014spin,zvyagin2014direct}, and theoretical calculations of the relevant exchange interactions, e.g., using {\it ab initio} \cite{jeschke2013first, riedl2019ab} or exact-diagonalization approaches \cite{zorko2015magnetic}, are required to determine the precise microscopic Hamiltonian of {\YCu} and thus to address the origin of magnetic ordering in this material in detail.

\vspace{0.5 cm}
The data that support the findings of this study are available in Durham University \cite{DurhamArxiv}.  

{\red 
\section{Acknowledgements}
We acknowledge the financial support by the Slovenian Research Agency under program No.~P1-0125.
M.G. is grateful to EPSRC (UK) for financial support through grant No.~EP/N024028/1.
This work is based on experiments performed at the Swiss spallation neutron source SINQ, Paul Scherrer Institute, Villigen, Switzerland.
}

%
\end{document}